\documentclass[conference,a4paper]{IEEEtran}
\usepackage{ifpdf}
\usepackage{cite}
\ifCLASSINFOpdf
   \usepackage[pdftex]{graphicx}
\else
\fi
\usepackage[cmex10]{amsmath}
\usepackage{url}
\usepackage{amssymb}
\usepackage[nolist]{acronym}
\usepackage{bm}
\usepackage{units}
\usepackage[noabbrev,capitalise]{cleveref}
\crefname{equation}{}{}
\usepackage{color}

\begin{document}
	% define acronyms
	\begin{acronym}
		\acro{PDF}{probability density function}
\acro{DoA}{direction-of-arrival}
\acrodefplural{DoA}[DoAs]{directions-of-arrival}
\acro{RSSI}{received signal strength indicator}
\acro{RHCP}{right hand circular polarisation}
\acro{WSN}{wireless sensor networks}
\acro{MUSIC}{multiple signal characterization}
\acro{ML}{maximum likelihood}
\acro{FIM}{Fisher information matrix}
\acro{CRB}{Cram\'{e}r-Rao bound}
\acro{ULA}{uniform linear array}
\acro{EMF}{electromagnetic field}
\acro{MEMS}{microelectromechanical system}
\acro{GNSS}{global navigation satellite system}
\acro{WSN}{wireless sensor networks}
\acro{SNR}{signal-to-noise ratio}
\acro{FOV}{field of view}
\acro{MMA}{multi-mode antenna}
\acro{MSE}{mean-square error}
\acro{RMSE}{root-mean-square error}
\acro{FRI}{finite-rate-of-innovation}
	\end{acronym}
	\title{Power-Based Direction-of-Arrival Estimation Using a Single Multi-Mode Antenna}

	\author{\IEEEauthorblockN{Robert P\"ohlmann, Siwei Zhang, Thomas Jost, Armin Dammann}
		\IEEEauthorblockA{\\German Aerospace Center (DLR)\\
			Institute of Communications and Navigation\\
			Oberpfaffenhofen, 82234 Wessling, Germany\\
			Email: \{Robert.Poehlmann, Siwei.Zhang, Thomas.Jost, Armin.Dammann\}@dlr.de}
		
		% this empty line is important!
	}

	
	%Telephone: (800) 555--1212, Fax: (888) 555--1212}
	
	\IEEEoverridecommandlockouts
	% copyright clearance code
	
	% switch off page numbering
	\pagenumbering{gobble}
	
	
	\maketitle
	
	\begin{abstract}
		%\boldmath
		Phased antenna arrays are widely used for \ac{DoA} estimation. For low-cost applications, signal power or \ac{RSSI} based approaches can be an alternative. However, they usually require multiple antennas, a single antenna that can be rotated, or switchable antenna beams. In this paper we show how a \ac{MMA} can be used for power-based \ac{DoA} estimation. Only a single \ac{MMA} is needed and neither rotation nor switching of antenna beams is required. We derive an estimation scheme as well as theoretical bounds and validate them through simulations. It is found that power-based \ac{DoA} estimation with an \ac{MMA} is feasible and accurate.
	\end{abstract}
	

	% \fi
	\IEEEpeerreviewmaketitle

	\acresetall
	\section{Introduction}\label{s:introduction}
	The state-of-the-art approach of \ac{DoA} estimation relies on signal phase differences (assuming narrowband signals) between the elements of an antenna array \cite{tuncer2009}. \ac{DoA} estimation methods are well known, however using an antenna array requires a complex receiver structure. For each array element, a separate receiver channel is required and the receiver channels have to be coherent and the system well calibrated. For low-cost applications, another possibility is to use signal power, i.e. \ac{RSSI} measurements for \ac{DoA} estimation. This requires knowledge of the antenna pattern and the possibility to cancel out or estimate the unknown path loss and transmit power of the signal.
	
	In this paper we focus on power-based \ac{DoA} estimation. Different techniques can be found in literature. One approach is to use an array of directional antennas pointing in different directions, see e.g. \cite{ash2004}. If only one antenna is available, an actuator can be used to rotate the antenna and thus obtain measurements from different angles. This approach may be used with either omnidirectional antennas having knowledge of the null in the pattern, see e.g. \cite{malajner2012}, or directional antennas, e.g. \cite{hood2011}. Instead of rotating the antenna, it is possible to make a controlled movement with the whole platform (e.g. a quadrocopter) \cite{isaacs2014}. Mechanical actuators have the drawback that they increase the power consumption of the system, possibly need maintenance and limit the update rate. Hence the authors in \cite{cidronali2010} avoid moving parts and propose a switched beam antenna. In general, high-resolution properties can be achieved with \ac{RSSI} measurements. In \cite{passafiume2015}, a variant of \ac{MUSIC} suitable for signal power measurements is presented. The authors in \cite{lie2010} apply methods known from \ac{FRI} sampling to obtain high-resolution. All methods in the literature that the authors are aware of have in common that they either use multiple antennas or a single antenna with some sort of rotational movement or beams that require switching.
	
	In contrast, this paper presents a power-based \ac{DoA} estimation scheme with a \ac{MMA}, which does not require any movement or switching of antenna beams. An \ac{MMA} is in fact a single antenna element. Compared to antenna arrays, it has the advantage of being more compact, which can be important for applications with size constraints. The remainder of this paper is organised as follows. In \cref{s:antenna}, the concept of the \ac{MMA} is introduced. \Cref{s:wavefield} quickly recapitulates the theoretical basis of this work termed wavefield modelling. The \ac{DoA} estimation scheme is then developed in \cref{s:doa}. The performance is evaluated in terms of theoretical bounds and simulations. \Cref{s:conclusion} concludes the paper.

	\section{Multi-Mode Antenna}\label{s:antenna}
	In this section we provide a brief introduction to \acp{MMA}. The concept of \acp{MMA} is based on the theory of characteristic modes \cite{garbacz1971,harrington1971a}. This theory is available for more than 40 years, with significant amount of attention over the last 15 years. Recently the theory is more popular among antenna designers and is now known to be a useful design aid \cite{lau2016}. The main idea is to express the surface current distribution of conducting bodies as a sum of orthogonal functions called characteristic modes. These modes are independent of the excitation, i.e. they are defined by the shape and the size of the conductor. It is possible to determine modes numerically for antennas of arbitrary shape. For electrically small conductors, few modes are sufficient to describe the antenna behaviour \cite{cabedo-fabres2007}. Hence, electrically small conductors are well suited for an application of the theory of characteristic modes. 
	
	The idea of \acp{MMA} is to excite different characteristic modes independently. The current distribution for the particular mode, found by the theory of characteristic modes, defines the locations for excitation. In general, excitation is possible by inductive coupling at the current maxima, or by capacitive coupling at the respective minima \cite{martens2011a}. The couplers belonging
	\newpage\noindent
	to one mode are then connected to one port of the antenna. Due to the different modes being orthogonal to each other, \acp{MMA} are able to provide sufficient isolation of the ports \cite{manteuffel2016}.
	
	Compared to classic antenna arrays \acp{MMA} are potentially more compact, which could be an advantage for applications with stringent size and weight constraints. The theory of characteristic modes provides a useful tool to design chassis antennas. For example, the case structure of a mobile handset device \cite{li2012} or the platform of an unmanned aerial vehicle (UAV) can be used as antenna \cite{chen2014}.
	
\begin{figure}
		\centering
		\includegraphics{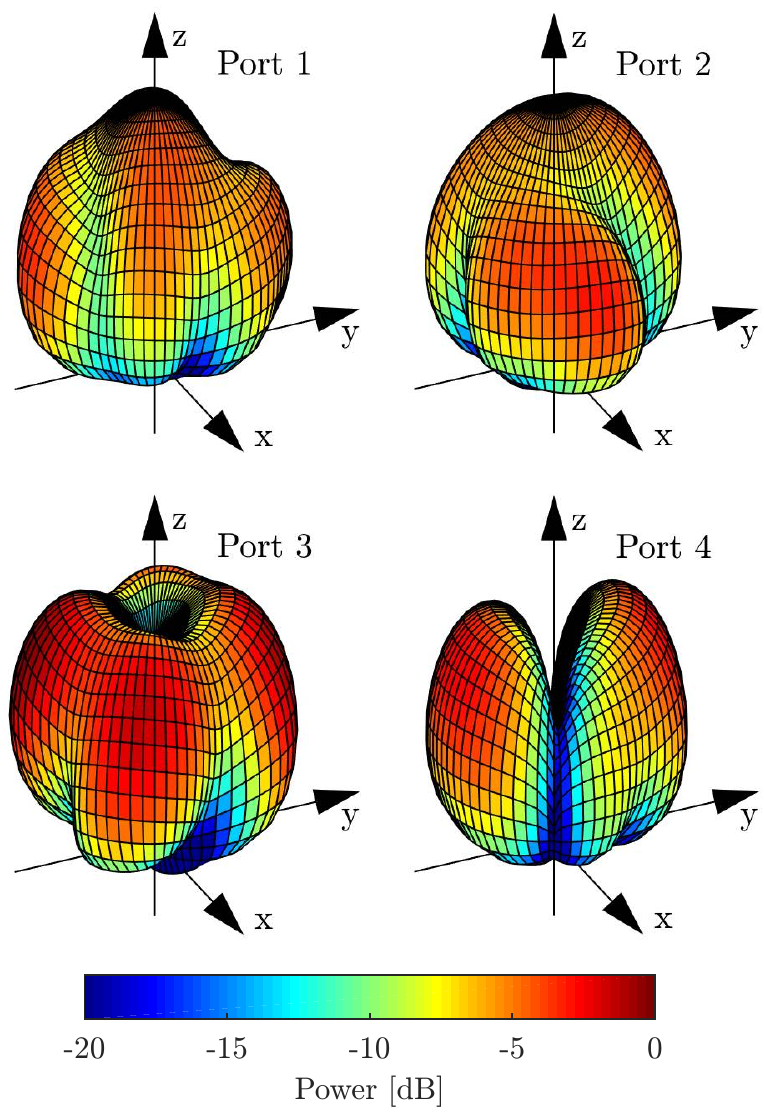}
		\caption{\ac{MMA} power patterns for \ac{RHCP}}
		\label{fig:mm3d}
	\end{figure}

\begin{figure}
		\centering
		\includegraphics{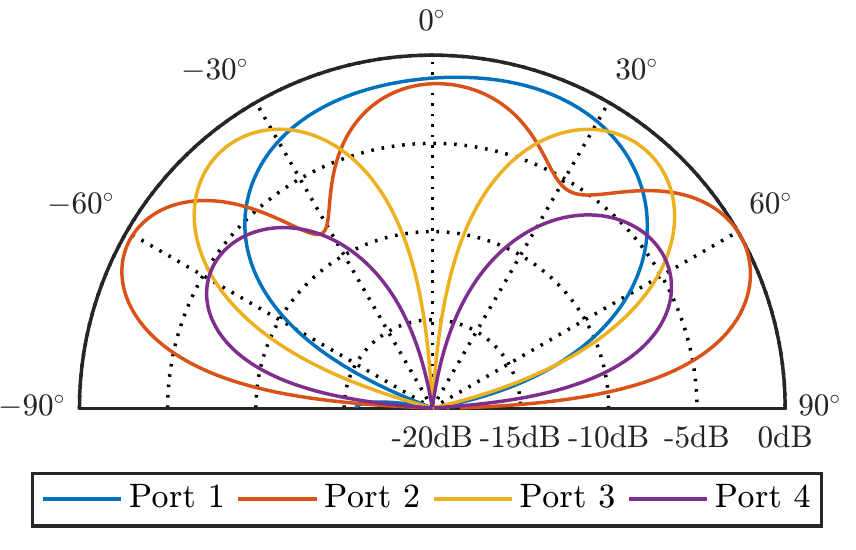}
		\caption{\ac{MMA} x-z plane power patterns for \ac{RHCP}}
		\label{fig:mm2d}
	\end{figure}

	To the authors knowledge, \acp{MMA} have so far been investigated only for communication applications, see e.g. \cite{hoeher2015}. The aim of this paper is to show how their potential can be used for \ac{DoA} estimation, enabling applications like localisation and orientation estimation. The antenna that we analyse in this paper has been proposed in \cite{manteuffel2016}. \Cref{fig:mm3d} shows the power pattern of this four port \ac{MMA} for \ac{RHCP}. The respective power pattern of the x-z plane is given in \cref{fig:mm2d}. Obviously the antenna gain strongly depends on the incident angle of the signal. Moreover, the antenna patterns differ significantly between the four ports. Therefore the received signal power could be used to estimate the \ac{DoA} of the signal. In the following, we assume that the polarisation of the received narrowband signal is purely \ac{RHCP}.

	\section{Wavefield Modelling}\label{s:wavefield}
\begin{figure}
		\centering
		\includegraphics{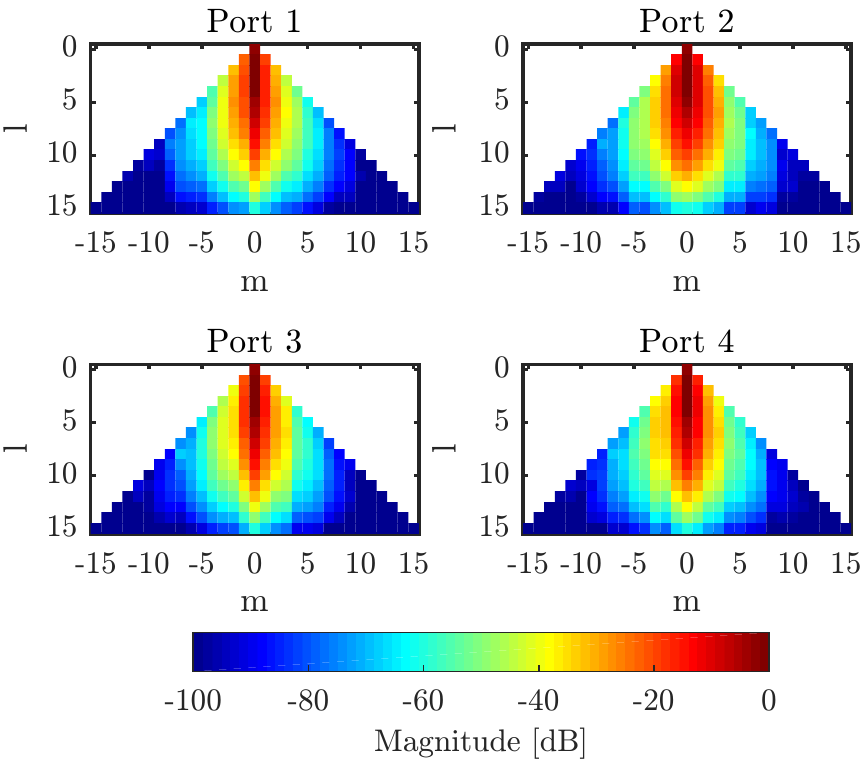}
		\caption{Magnitude of spherical harmonic coefficients, indices $l$ and $m$ are degree and order of the spherical harmonics in \cref{eq:sh}}
		\label{fig:SH_mag}
	\end{figure}

	The \ac{MMA} is described in terms of spatial samples of the antenna power pattern obtained by \ac{EMF} simulation or calibration measurements. Since that representation might be relatively sparse, an interpolation strategy is needed, where we apply wavefield modelling and manifold separation \cite{doron1994}. In general, the manifold of the \ac{MMA} is defined by $\theta \in [-\pi,\pi)$ for 2D and $\theta \in [0,\pi], \phi \in [0,2\pi)$ for the 3D case, where $\theta$ is the co-elevation and $\phi$ the azimuth. If a function is square integrable on this manifold, then it can be expanded in terms of an orthonormal basis. Hence the power pattern of an \ac{MMA} with $M$ ports, $\bm{g}(\theta,\phi) \in \mathbb{R}^{M \times 1}$, can be decomposed \cite{doron1994}, such that
	\begin{equation}\label{eq:ms}
	\bm{g}(\theta,\phi) = \bm{G} \, \bm{\Psi}(\theta,\phi).
	\end{equation}
	The matrix $\bm{G} \in \mathbb{R}^{M \times N}$ is called sampling matrix and $\bm{\Psi}(\theta,\phi) \in \mathbb{R}^{N \times 1}$ is the basis vector, with $N$ being the order of the basis. For 2D, the Fourier functions,
	\begin{equation}\label{eq:Psitheta}
	\bm{\Psi}(\theta) = \frac{1}{\sqrt{2\pi}}e^{j\bm{n}\theta}, \: \bm{n}=[-N,...,N],
	\end{equation}
	can be used as a basis. Then  $\bm{G}_{m,N/2+1:N} = \bm{G}_{m,1:N/2}^*$, because $\bm{g}(\theta)$ must be real valued. For 3D we use the real spherical harmonic functions,
	\begin{equation}\label{eq:sh}
	Y_l^m(\theta,\phi) = 
	\begin{cases}
	\sqrt{2} N_l^m \cos(m\phi) P_l^m(\cos(\theta)) & m>0 \\
	N_l^0 P_l^m(\cos(\theta)) & m=0 \\
	\sqrt{2} N_l^{|m|} \sin(|m|\phi) P_l^{|m|}(\cos(\theta)) & m<0, \\
	\end{cases}
	\end{equation}
	with degree $l = 0,...,L$ and order $m = -l,...,l$. $P_l^m(.)$ is the associated Legendre polynomial with degree $l$ and order $m$. The normalization factor $N_l^m$ is given by
	\begin{equation}
	N_l^m = \sqrt{\frac{2l+1}{4\pi}\frac{(l-m)!}{(l+m)!}}.
	\end{equation}
	Using the enumeration $n=(l+1)l+m$, we can form an orthonormal basis,
	\begin{equation}
	\bm{\Psi}(\theta,\phi) = \bm{Y}_{\bm{n}}(\theta,\phi), \: \bm{n}=[0,...,N].
	\end{equation}
	It is known that the magnitude of $\bm{G}$ decays superexponentially for $n \rightarrow \infty$ beyond $n=kr$, with $k$ being the angular wavenumber and $r$ the radius of the smallest sphere enclosing the antenna \cite{doron1994}. \Cref{fig:SH_mag} shows the magnitude of $\bm{G}$, i.e. the spherical harmonic coefficients, for the \ac{MMA}. As can be seen, most of the energy is contained in the low order coefficients. Hence the expansion can be safely truncated at a certain order. As \cref{eq:ms} is linear, having enough spatial samples of the antenna power pattern $\bm{\tilde{g}}(\theta,\phi)$ available from calibration measurements or \ac{EMF} simulation, it is straightforward to determine the sampling matrix $\bm{G}$ for a given basis $\bm{\Psi}(\theta,\phi)$.
	
	Due to the assumption of narrowband and pure polarisation, it is sufficient to consider a scalar wavefield. Nevertheless, an extension to vector fields is possible \cite{costa2012}. Having a suitable model of the antenna at hand, we will now take a look at power-based \ac{DoA} estimation.

	\section{Direction-of-Arrival Estimation}\label{s:doa}
	For simplicity, the derivations and simulations in \cref{ss:signal,ss:ML,ss:CRB,ss:simulation} focus on the 2D case, i.e. only a single angle $\theta$ is to be estimated. The extension to 3D and two angles of arrival follows in \cref{ss:3D}.
	
	\subsection{Signal Model}\label{ss:signal}
	The received, sampled signal $r_m(k)$ at port $m$ of the \ac{MMA} is given by
	\begin{equation}
	r_m(k) = a_m(\theta) s(k) + n_m(k)
	\end{equation}
	where $k$ is the sample index, $a_m(\theta)$ is the attenuation caused by the antenna, $s(k)$ is the transmitted signal as it arrives at the receive antenna and $n_m(k) \sim \mathcal{CN}(0, \sigma^2)$ is white circular symmetric normal distributed noise with variance $\sigma^2$. Assuming stationarity, the time-averaged received signal power over $K$ samples in time can be calculated by
	%\begin{equation}
	%\begin{split}
	%	E[P_{\text{r},m}] &= \frac{1}{K} \sum_{k=1}^{K} |r_m(k)|^2 \\
	%	&= \underbrace{|a_m(\theta)|^2}_{g_m(\theta)} \frac{1}{K} \sum_{k=1}^{K} |s(k)|^2 + \frac{1}{K} \sum_{k=1}^{K} |n_m(k)|^2 \\
	%	&= g_m(\theta) P_{\text{s}} + P_{n,m}
	%\end{split}
	%\end{equation}
	%\begin{equation}\label{eq:Prm}
	%\begin{split}
	%	\text{E}[P_{\text{r},m}] &= \text{E}\left[ \frac{1}{K} \sum_{k=1}^{K} |r_m(k)|^2 \right] \\
	%	&= \underbrace{|a_m(\theta)|^2}_{g_m(\theta)} \frac{1}{K} \sum_{k=1}^{K} |s(k)|^2 + \text{E}\left[ \frac{1}{K} \sum_{k=1}^{K} |n_m(k)|^2 \right] \\
	%	&= g_m(\theta) P_{\text{s}} + \sigma^2
	%\end{split}
	%\end{equation}
	\begin{equation}\label{eq:Prm}
	\begin{split}
	\text{E}[P_{\mathrm{r},m}] &= \text{E}\left[ \frac{1}{K} \sum_{k=1}^{K} |r_m(k)|^2 \right] \\
	&= \underbrace{|a_m(\theta,\phi)|^2}_{g_m(\theta,\phi)} \frac{1}{K} \sum_{k=1}^{K} |s(k)|^2 + \text{E}\left[ \frac{1}{K} \sum_{k=1}^{K} |n_m(k)|^2 \right] \\
	&= g_m(\theta,\phi) P_\mathrm{s} + \sigma^2,
	\end{split}
	\end{equation}
	with the antenna power pattern $g_m(\theta)$ and the signal power $P_{\text{s}}$. We follow an \ac{RSSI} based approach, hence only power measurements are available to the receiver. The antenna power patterns, $\bm{g}(\theta) = [g_1(\theta),...,g_M(\theta)]$, are normalized such that $\max \bm{g}(\theta) = 1$. The SNR for $g_m(\theta)=1$ is then given by
	\begin{equation}
	\text{SNR} = \frac{P_{\text{s}}}{\sigma^2}.
	\end{equation}
	Defining $r_{m,\text{r}}(k) = \text{Re}[r_m(k)]$ and $r_{m,\text{i}}(k) = \text{Im}[r_m(k)]$, the sum of the squared magnitude of the received signal,
	\begin{equation}
	S_{\text{r},m} = \sum_{k=1}^{K} |r_m(k)|^2 = \sum_{k=1}^{K} r_{m,\text{r}}^2(k) + r_{m,\text{i}}^2(k) \sim \chi^2(2K,\lambda,\sigma^2/2)
	\end{equation}
	follows a noncentral $\chi^2$ distribution \cite{proakis2008} with $2K$ degrees of freedom and noncentrality parameter
	%\begin{equation}\label{eq:lambda}
	%	\lambda = \sum_{k=1}^{K} \left( E[r_{m,\text{r}}(k)]^2 + E[r_{m,\text{i}}(k)]^2 \right) = K g_m(\theta) P_{\text{s}}.
	%\end{equation}
	\begin{equation}\label{eq:lambda}
	\begin{split}
	\lambda &= \sum_{k=1}^{K} \left( E[r_{m,\text{r}}(k)]^2 + E[r_{m,\text{i}}(k)]^2 \right) \\
	&= \sum_{k=1}^{K} (a_m(\theta))^2 (\text{Re}[s(k)]^2 + \text{Im}[s(k)]^2) \\
	&= \sum_{k=1}^{K} g_m(\theta)  |s(k)|^2 \\
	&= K g_m(\theta) P_{\text{s}}.
	\end{split}
	\end{equation}
	Its \ac{PDF} is given by
	\begin{equation}
	p_{S_{\text{r},m}}(x) = \frac{1}{\sigma^2} \left(\frac{x}{\lambda}\right)^{\frac{K-1}{2}} e^{-\frac{\lambda+x}{\sigma^2}} I_{K-1}\left(\frac{2\sqrt{\lambda x}}{\sigma^2}\right),
	\end{equation}
	with the modified Bessel function of the first kind $I_\nu(.)$. Since $P_{\text{r},m}$ is just a scaled version of that, its distribution can be obtained by transformation $p_{P_{\text{r},m}}(x)= K  p_{S_{\text{r},m}}(Kx)$.
	%\begin{equation}
	%p(P_{\text{r},m}) = \frac{K}{2\sigma^2} \left(\frac{KP_{\text{r},m}}{\lambda}\right)^{\frac{K-1}{2}} e^{-\frac{\lambda+KP_{\text{r},m}}{2\sigma^2}} I_{K-1}\left(\frac{\sqrt{\lambda K P_{\text{r},m}}}{\sigma^2}\right)
	%\end{equation}
	Inserting \cref{eq:lambda}, we obtain
	\begin{multline}\label{eq:pdf}
	p_{P_{\text{r},m}}(x) = \frac{K}{\sigma^2} \left(\frac{x}{g_m(\theta)P_{\text{s}}}\right)^{\frac{K-1}{2}} e^{-\frac{K(g_m(\theta)P_{\text{s}}+x)}{\sigma^2}}\\ I_{K-1}\left(\frac{2K\sqrt{g_m(\theta) P_{\text{s}}  x}}{\sigma^2}\right).
	\end{multline}
	The mean and variance are given by
	\begin{equation}\label{eq:mu}
	\begin{split}
	\tilde{\mu}_m &= \text{E}[P_{\text{r},m}] = K^{-1} \, \text{E}[S_{r,m}] \\
	&= K^{-1} (K\sigma^2+\lambda) = g_m(\theta) P_{\text{s}} + \sigma^2,
	\end{split}
	\end{equation}
	\begin{equation}\label{eq:sigma}
	\begin{split}
	\tilde{\sigma}_m^2 &= \text{VAR}[P_{\text{r},m}] = K^{-2} \, \text{VAR}[S_{r,m}] \\
	&= K^{-2} (K\sigma^4+2\sigma^2\lambda) \\ &= K^{-1}(\sigma^4+2g_m(\theta)P_{\text{s}}\sigma^2).
	\end{split}
	\end{equation}
	For large $\lambda$ or large $K$, \cref{eq:pdf} is approximately Gaussian distributed $P_{\text{r},m} \sim \mathcal{N}(\tilde{\mu}_m, \tilde{\sigma}_m^2)$.
	
	\subsection{ML Estimator}\label{ss:ML}
	The parameters to be estimated are given by
	\begin{equation}
	\bm{\Gamma} = [\theta, P_{\text{s}}, \sigma^2].
	\end{equation}
	We consider the case of unknown signal power $P_{\text{s}}$ and noise variance $\sigma^2$. Using the Gaussian approximation, i.e. $K$ is large, the log-likelihood function for the signal power measurements $\bm{P_\text{r}} = [P_{\text{r},1}, ..., P_{\text{r},M}]$ can be written as
	\begin{multline}\label{eq:gausspdf}
	\ln p(\bm{P_\text{r}}; \bm{\Gamma}) = \sum_{m=1}^{M} -\frac{1}{2}\ln(2\pi\tilde{\sigma}_m^2)\\ -\frac{1}{2\tilde{\sigma}_m^2} \left( P_{\text{r},m} - (g_m(\theta)P_{\text{s}}+\sigma^2 )\right)^2.
	\end{multline}
	The corresponding \ac{ML} estimator can then be derived as
	\begin{equation}\label{eq:ml}
	\begin{split}
	\bm{\hat{\Gamma}_\text{ML}} &= \arg \max_{\bm{\Gamma}} \ln p(\bm{P_r}; \bm{\Gamma}) \\
	&=\arg \min_{\bm{\Gamma}} \sum_{m=1}^{M} \ln \left( 2 \pi K^{-1} (\sigma^4 + 2 g_m(\theta) P_{\text{s}} \sigma^2) \right) \\
	&\quad + \frac{K(P_{\text{r},m} - P_{\text{s}} g_m(\theta) - \sigma^2)^2}{\sigma^4 + 2 P_{\text{s}} g_m(\theta) \sigma^2}.
	\end{split}
	\end{equation}
	A simplified version of the estimator,
	\begin{equation}\label{eq:mlapprox}
	\bm{\hat{\Gamma}_\text{S}} =
	\arg \min_{\bm{\Gamma}} \sum_{m=1}^{M} \frac{(P_{\text{r},m} - P_{\text{s}} g_m(\theta) - \sigma^2)^2}{P_{\text{s}} g_m(\theta) + 2\sigma^2},
	\end{equation}
	can be obtained by neglecting the logarithmic term.

	\subsection{CRB Derivation}\label{ss:CRB}
	A lower bound on the variance of any unbiased estimator is the \ac{CRB} \cite{kay1993}. For a given set of unknowns $\bm{\Gamma} = [\theta, P_{\text{s}}, \sigma^2]$, it is defined as the inverse of the \ac{FIM} $ \bm{I}(\bm{\Gamma})$,
	\begin{equation}
	\text{VAR}[\bm{\hat{\Gamma}}] \geq \bm{I}(\bm{\Gamma})^{-1}.
	\end{equation}
	Following the Gaussian assumption in \cref{eq:gausspdf}, the elements of the \ac{FIM} $\bm{I}(\bm{\Gamma}) \in \mathbb{R}^{3 \times 3}$ can be calculated as \cite{kay1993}
	\begin{equation}
	[\bm{I}(\Gamma)]_{i,j} = \sum_{m=1}^{M} \frac{1}{\tilde{\sigma}_m^2} \frac{\partial\tilde{\mu}_m}{\partial\Gamma_i} \frac{\partial\tilde{\mu}_m}{\partial\Gamma_j} + \frac{1}{2\tilde{\sigma}_m^4} \frac{\partial\tilde{\sigma}_m^2}{\partial\Gamma_i} \frac{\partial\tilde{\sigma}_m^2}{\partial\Gamma_j}.
	\end{equation}
	Calculation of the partial derivatives of \cref{eq:mu,eq:sigma} requires the derivative of \cref{eq:ms,eq:Psitheta}, which is given by
	\begin{equation}
	\frac{\partial \bm{g}(\theta)}{\partial \theta} = \bm{G} \frac{j\bm{n}}{\sqrt{2\pi}}e^{j\bm{n}\theta}, \: \bm{n}=[-N,...,N].
	\end{equation}
	Finally we obtain the \ac{CRB} for the estimation of $\theta$,
	\begin{equation}
	\text{VAR}[\theta] \geq [\bm{I}(\bm{\Gamma})^{-1}]_{1,1} = \text{CRB}(\theta).
	\end{equation}

	\subsection{Simulation Results}\label{ss:simulation}
\begin{figure}
		\centering
		\includegraphics{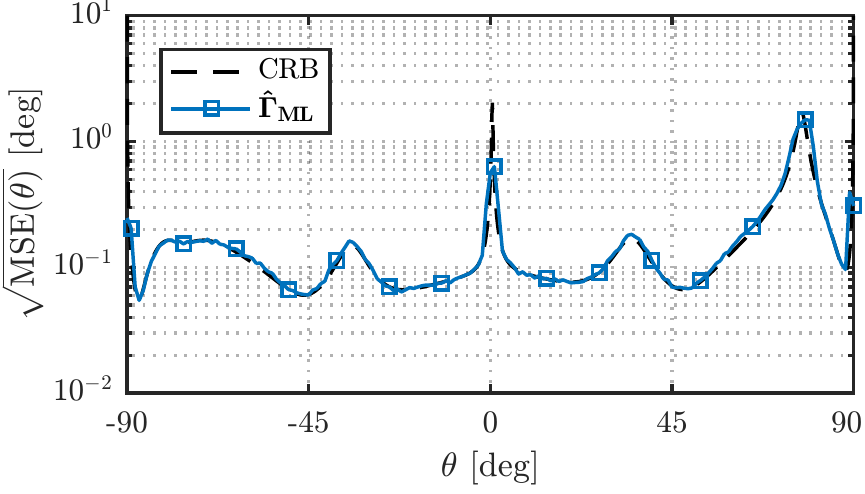}
		\caption{\ac{RMSE} of the simulation and \ac{CRB} for $\text{SNR}=\unit[20]{dB}$}
		\label{fig:MSE_theta}
	\end{figure}

	In order to evaluate the feasibility of the proposed \ac{DoA} estimation approach, we have used \ac{EMF} simulation data of the \ac{MMA} prototype, visualized in \cref{fig:mm2d}. The simulated \ac{RMSE} and the \ac{CRB} depending on $\theta$ are shown in \cref{fig:MSE_theta}. The plot indicates that the achievable accuracy depends on the incident angle, with a \ac{RMSE} spread of more than one order of magnitude over the manifold. Nevertheless, the \ac{ML} estimator is able to attain the \ac{CRB} for high \ac{SNR}.
	\begin{figure}
		\centering
		\includegraphics{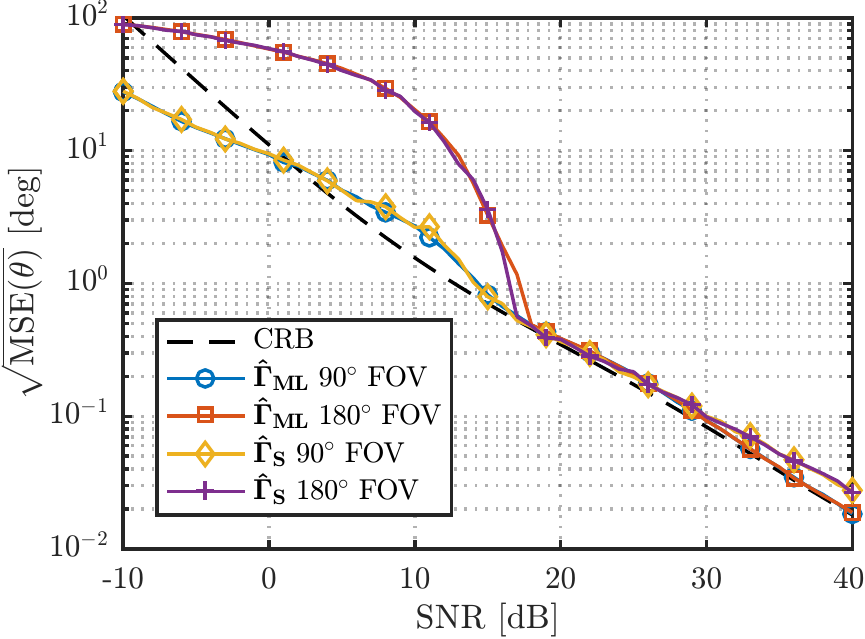}
		\caption{Simulated \ac{RMSE} of \ac{ML} and simplified estimator and \ac{CRB} vs. \ac{SNR} for full ($180^\circ$) and limited ($90^\circ$) FOV}
		\label{fig:MSE_SNR}
	\end{figure}
	
	In \cref{fig:MSE_SNR} the mean over the manifold is plotted in dependence of the \ac{SNR} present at the receiver. We have seen earlier, in \cref{fig:mm2d}, that the antenna patterns are relatively symmetric around $0^\circ$. For that reason, we compare 
	the performance with a limited \ac{FOV} of $90^\circ$ to the full \ac{FOV} of $180^\circ$. It can be seen 
	that for $\text{SNR} \geq \unit[18]{dB}$, the \ac{ML} estimator $\bm{\hat{\Gamma}_\text{ML}}$ defined in \cref{eq:ml} is efficient, i.e. it attains the \ac{CRB}. For lower SNR, the \ac{RMSE} is significantly bigger for $180^\circ$ due to ambiguities caused by the antenna pattern. The \ac{CRB} is calculated independent of the a-priori information regarding \ac{FOV} limitation, hence it is not an accurate lower bound in the low \ac{SNR} region. Finally the plot indicates that the simplified estimator $\bm{\hat{\Gamma}_\text{S}}$ given by \cref{eq:mlapprox} is sufficiently accurate. Only for high \ac{SNR}, a slight increase in \ac{RMSE} compared to the \ac{ML} estimator is visible.

	\subsection{Extension to 3D}\label{ss:3D}
	\begin{figure}[t]
		\centering
		\includegraphics{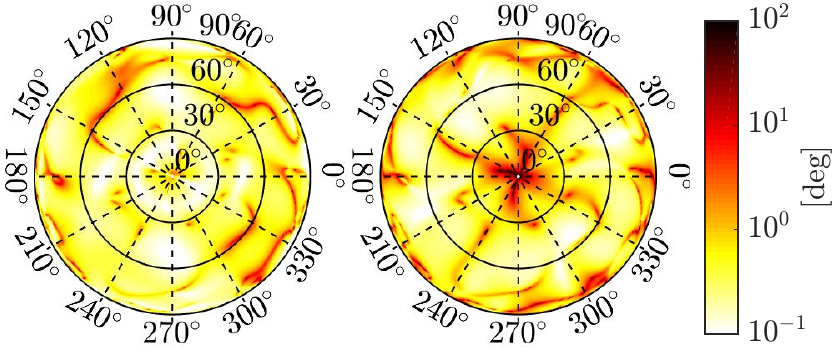}
		\caption{$\sqrt{CRB(\theta)}$ (left) and $\sqrt{CRB(\phi)}$ (right) for $\text{SNR} = \unit[20]{dB}$}
		\label{fig:CRB_3D_20dB}
	\end{figure}
	\begin{figure}[t]
		\centering
		\includegraphics{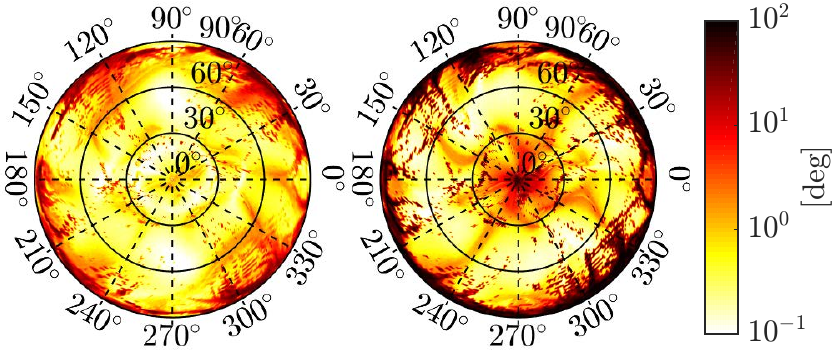}
		\caption{$\sqrt{MSE(\theta)}$ (left) and $\sqrt{MSE(\phi)}$ (right) for $\text{SNR} = \unit[20]{dB}$}
		\label{fig:MSE_3D_20dB}
	\end{figure}
	\begin{figure}[t]
		\centering
		\includegraphics{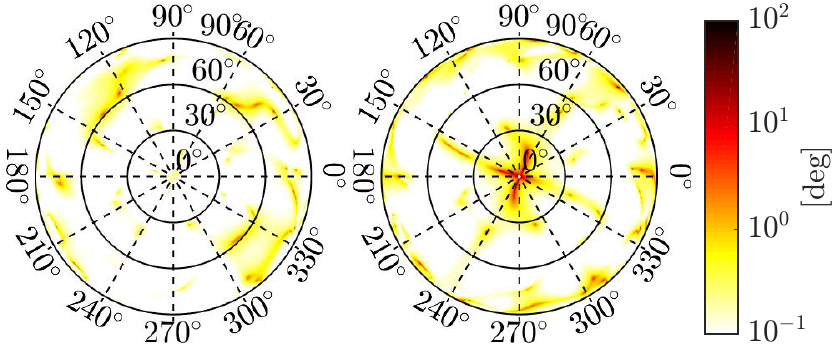}
		\caption{$\sqrt{CRB(\theta)}$ (left) and $\sqrt{CRB(\phi)}$ (right) for $\text{SNR} = \unit[30]{dB}$}
		\label{fig:CRB_3D_30dB}
	\end{figure}
	\begin{figure}[t]
		\centering
		\includegraphics{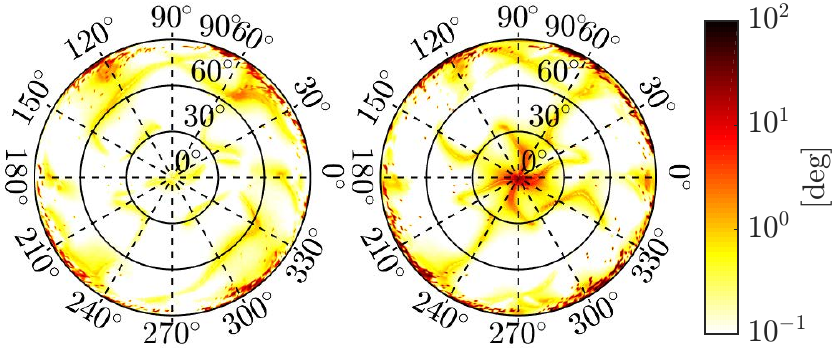}
		\caption{$\sqrt{MSE(\theta)}$ (left) and $\sqrt{MSE(\phi)}$ (right) for $\text{SNR} = \unit[30]{dB}$}
		\label{fig:MSE_3D_30dB}
	\end{figure}
	Having studied the performance in 2D, we now extend the proposed \ac{DoA} estimation scheme to the more practical case of 3D, i.e. two unknown angles of arrival. The extension of the signal model to 3D,
	\begin{equation}
	r_m(k) = a_m(\theta,\phi) s(k) + n_m(k),
	\end{equation}
	is straight forward. One more parameter, $\phi$, has to be estimated, so we have $\bm{\Gamma} = [\theta, \phi, P_{\text{s}}, \sigma^2]$. The \ac{FIM} then grows to $I(\bm{\Gamma}) \in \mathbb{R}^{4 \times 4}$. Applying wavefield modelling with spherical harmonics as described in \cref{s:wavefield}, the partial derivatives of the antenna power pattern are
	\begin{subequations}
		\begin{align}
		&\frac{\partial \bm{g}(\theta,\phi)}{\partial\theta} = \bm{G} \frac{\partial}{\partial\theta} Y_n(\theta, \phi), \\
		&\frac{\partial \bm{g}(\theta,\phi)}{\partial\phi} = \bm{G} \frac{\partial}{\partial\phi} Y_n(\theta, \phi),
		\end{align}
	\end{subequations}
	using the enumeration $n=(l+1)l+m$. The partial derivative of \cref{eq:sh} with respect to $\theta$ (for $\theta \neq 0$) is given by
	\begin{equation}
	\frac{\partial}{\partial\theta} Y_l^m(\theta,\phi) =
	\begin{cases}
	\sqrt{2} N_l^m \cos(m\phi) \frac{\partial P_l^m(\cos(\theta))}{\partial\theta} & m>0 \\
	N_l^0 \frac{\partial P_l^m(\cos(\theta))}{\partial\theta}  & m=0 \\
	\sqrt{2} N_l^{|m|} \sin(|m|\phi) \frac{\partial P_l^{|m|}(\cos(\theta))}{\partial\theta} & m<0. \\
	\end{cases}
	\end{equation}
	The derivative of the Legendre polynomial $P_l^m(\cos(\theta))$,
	\begin{equation}
	\frac{\partial P_l^m(\cos(\theta))}{\partial\theta} = \frac{1+l-m}{\sin(\theta)}P_{l+1}^m(cos(\theta)) - \frac{l+1}{\tan(\theta)}P_l^m(\cos(\theta)),
	\end{equation}
	can be calculated with the help of \cite{olver2010}. The corresponding partial derivative of \cref{eq:sh} with respect to $\phi$ is given by
	\begin{equation}
	\frac{\partial}{\partial\phi} Y_l^m(\theta,\phi) =
	\begin{cases}
	\sqrt{2} N_l^m (-m)\sin(m\phi) P_l^m(\cos(\theta)) & m>0 \\
	0 & m=0 \\
	\sqrt{2} N_l^{|m|} (-m)\cos(m\phi) P_l^{|m|}(\cos(\theta)) & m<0. \\
	\end{cases}
	\end{equation}
	Finally we obtain the \acp{CRB} for the estimation of $\theta$ and $\phi$ in the 3D case,
	\begin{subequations}
		\begin{equation}
		\text{VAR}[\hat{\theta}] \geq [\bm{I}(\bm{\Gamma})^{-1}]_{1,1} = \text{CRB}(\theta),
		\end{equation}
		\begin{equation}
		\text{VAR}[\hat{\phi}] \geq [\bm{I}(\bm{\Gamma})^{-1}]_{2,2} = \text{CRB}(\phi).
		\end{equation}
	\end{subequations}
	
	In order to confirm the expected \ac{DoA} estimation performance, we performed simulations for the 3D case. \Cref{fig:CRB_3D_20dB} shows the \ac{CRB} in $\theta$- and $\phi$-domain for $\text{SNR} = \unit[20]{dB}$. The corresponding  simulation result can be seen in \cref{fig:MSE_3D_20dB}. Apparently, the \ac{CRB} is not attained on the whole manifold. Especially for low elevations, an excessive estimation error can be observed. Taking another look at the antenna power pattern in \cref{fig:mm3d}, it is obvious that at low elevations the antenna gain is very small. This leads to a degradation of the \ac{DoA} estimation performance. For mid and high elevations, the \ac{CRB} is usually attained, except for a few directions which appear as dark dots in \cref{fig:MSE_3D_20dB}. This is most likely caused by estimation ambiguities due to the symmetry of the antenna pattern. In practice a-priori information, i.e. a rough knowledge of the direction, is often available, which may help to solve the ambiguity. Next we take a look at \cref{fig:CRB_3D_30dB,fig:MSE_3D_30dB} showing the \ac{CRB} and simulation \ac{RMSE} at $\text{SNR} = \unit[30]{dB}$. To allow comparison, \cref{fig:CRB_3D_20dB,fig:MSE_3D_20dB,fig:CRB_3D_30dB,fig:MSE_3D_30dB} use the same scaling. It can be seen that by increasing the \ac{SNR} by $\unit[10]{dB}$, both perturbing effects, i.e. low gain at low elevation angles and estimation ambiguities, are strongly reduced. Only for low elevations, an increased error can still be observed. The higher \ac{RMSE} at high elevation for the $\phi$-domain is less problematic, since at the pole this translates to a smaller directional error.

	\section{Conclusion}\label{s:conclusion}
	A suitable model for \acp{MMA}, building on the concept of wavefield modelling, was introduced. Based on that model, a power-based maximum likelihood \ac{DoA} estimation scheme has been introduced. Simulations have shown that in general, power-based \ac{DoA} estimation with \acp{MMA} is feasible. However, for certain incident angles, ambiguities in the antenna pattern cause an increased estimation variance. For the investigated \ac{MMA} prototype, low elevations are problematic due to low antenna gain. Altogether it can be concluded that power-based \ac{DoA} estimation is possible with \acp{MMA}, but for accurate estimates a relatively high \ac{SNR} is required. Further work will be performed by using signal polarisations as well as investigations into coherent receivers, that are able to obtain phase information from the different ports of the \ac{MMA}.
	

	\section*{Acknowledgment}
	This work has been funded by the German Research Foundation (DFG) under contract no. HO 2226/17-1. The authors are grateful for the constructive cooperation with Sami Alkubti Almasri, Niklas Doose and Prof. Peter A. Hoeher from the University of Kiel.
	
	Also, the authors would like to thank Prof. Dirk Manteuffel and his team for providing the antenna pattern of the \ac{MMA} prototype investigated in this paper.

	\bibliography{bibliography}

\begin{thebibliography}{10}
\providecommand{\url}[1]{#1}
\csname url@samestyle\endcsname
\providecommand{\newblock}{\relax}
\providecommand{\bibinfo}[2]{#2}
\providecommand{\BIBentrySTDinterwordspacing}{\spaceskip=0pt\relax}
\providecommand{\BIBentryALTinterwordstretchfactor}{4}
\providecommand{\BIBentryALTinterwordspacing}{\spaceskip=\fontdimen2\font plus
\BIBentryALTinterwordstretchfactor\fontdimen3\font minus
  \fontdimen4\font\relax}
\providecommand{\BIBforeignlanguage}[2]{{%
\expandafter\ifx\csname l@#1\endcsname\relax
\typeout{** WARNING: IEEEtran.bst: No hyphenation pattern has been}%
\typeout{** loaded for the language `#1'. Using the pattern for}%
\typeout{** the default language instead.}%
\else
\language=\csname l@#1\endcsname
\fi
#2}}
\providecommand{\BIBdecl}{\relax}
\BIBdecl

\bibitem{tuncer2009}
T.~E. Tuncer and B.~Friedlander, \emph{\BIBforeignlanguage{en}{Classical and
  {{Modern Direction}}-of-{{Arrival Estimation}}}}.\hskip 1em plus 0.5em minus
  0.4em\relax {Academic Press}, Jul. 2009.

\bibitem{ash2004}
J.~Ash and L.~Potter, ``Sensor network localization via received signal
  strength measurements with directional antennas,'' in \emph{Proceedings of
  the 2004 {{Allerton Conference}} on {{Communication}}, {{Control}}, and
  {{Computing}}}, 2004, pp. 1861--1870.

\bibitem{malajner2012}
M.~Malajner, P.~Planinsic, and D.~Gleich, ``Angle of {{Arrival Estimation Using
  RSSI}} and {{Omnidirectional Rotatable Antennas}},'' \emph{IEEE Sensors
  Journal}, vol.~12, no.~6, pp. 1950--1957, Jun. 2012.

\bibitem{hood2011}
B.~N. Hood and P.~Barooah, ``Estimating {{DoA From Radio}}-{{Frequency RSSI
  Measurements Using}} an {{Actuated Reflector}},'' \emph{IEEE Sensors
  Journal}, vol.~11, no.~2, pp. 413--417, Feb. 2011.

\bibitem{isaacs2014}
J.~T. Isaacs, F.~Quitin, L.~R.~G. Carrillo, U.~Madhow, and J.~P. Hespanha,
  ``Quadrotor control for {{RF}} source localization and tracking,'' in
  \emph{2014 {{International Conference}} on {{Unmanned Aircraft Systems}}
  ({{ICUAS}})}, May 2014, pp. 244--252.

\bibitem{cidronali2010}
A.~Cidronali, S.~Maddio, G.~Giorgetti, and G.~Manes, ``Analysis and
  {{Performance}} of a {{Smart Antenna}} for 2.45-{{GHz Single}}-{{Anchor
  Indoor Positioning}},'' \emph{IEEE Transactions on Microwave Theory and
  Techniques}, vol.~58, no.~1, pp. 21--31, Jan. 2010.

\bibitem{passafiume2015}
M.~Passafiume, S.~Maddio, A.~Cidronali, and G.~Manes, ``{{MUSIC}} algorithm for
  {{RSSI}}-based {{DoA}} estimation on standard {{IEEE}} 802.11/802.15. x
  systems,'' \emph{WSEAS Trans. Signal Process.}, vol.~11, pp. 58--68, 2015.

\bibitem{lie2010}
J.~P. Lie, T.~Blu, and C.~M.~S. See, ``Single {{Antenna Power Measurements
  Based Direction Finding}},'' \emph{IEEE Transactions on Signal Processing},
  vol.~58, no.~11, pp. 5682--5692, Nov. 2010.

\bibitem{garbacz1971}
R.~Garbacz and R.~Turpin, ``A generalized expansion for radiated and scattered
  fields,'' \emph{IEEE Transactions on Antennas and Propagation}, vol.~19,
  no.~3, pp. 348--358, 1971.

\bibitem{harrington1971a}
R.~Harrington and J.~Mautz, ``Theory of characteristic modes for conducting
  bodies,'' \emph{IEEE Transactions on Antennas and Propagation}, vol.~19,
  no.~5, pp. 622--628, 1971.

\bibitem{lau2016}
B.~K. Lau, D.~Manteuffel, H.~Arai, and S.~V. Hum, ``Guest {{Editorial Theory}}
  and {{Applications}} of {{Characteristic Modes}},'' \emph{IEEE Transactions
  on Antennas and Propagation}, vol.~64, no.~7, pp. 2590--2594, Jul. 2016.

\bibitem{cabedo-fabres2007}
M.~Cabedo-Fabres, E.~Antonino-Daviu, A.~Valero-Nogueira, and M.~F. Bataller,
  ``The theory of characteristic modes revisited: {{A}} contribution to the
  design of antennas for modern applications,'' \emph{IEEE Antennas and
  Propagation Magazine}, vol.~49, no.~5, pp. 52--68, 2007.

\bibitem{martens2011a}
R.~Martens, E.~Safin, and D.~Manteuffel, ``Inductive and capacitive excitation
  of the characteristic modes of small terminals,'' in \emph{Antennas and
  {{Propagation Conference}} ({{LAPC}}), 2011 {{Loughborough}}}.\hskip 1em plus
  0.5em minus 0.4em\relax {IEEE}, 2011, pp. 1--4.

\bibitem{manteuffel2016}
D.~Manteuffel and R.~Martens, ``Compact {{Multimode Multielement Antenna}} for
  {{Indoor UWB Massive MIMO}},'' \emph{IEEE Transactions on Antennas and
  Propagation}, vol.~64, no.~7, pp. 2689--2697, Jul. 2016.

\bibitem{li2012}
H.~Li, Y.~Tan, B.~K. Lau, Z.~Ying, and S.~He, ``Characteristic {{Mode Based
  Tradeoff Analysis}} of {{Antenna}}-{{Chassis Interactions}} for {{Multiple
  Antenna Terminals}},'' \emph{IEEE Transactions on Antennas and Propagation},
  vol.~60, no.~2, pp. 490--502, Feb. 2012.

\bibitem{chen2014}
Y.~Chen and C.~F. Wang, ``Electrically {{Small UAV Antenna Design Using
  Characteristic Modes}},'' \emph{IEEE Transactions on Antennas and
  Propagation}, vol.~62, no.~2, pp. 535--545, Feb. 2014.

\bibitem{hoeher2015}
P.~A. Hoeher and N.~Doose, ``\BIBforeignlanguage{en}{A massive {{MIMO}}
  terminal concept based on small-size multi-mode antennas},''
  \emph{\BIBforeignlanguage{en}{Transactions on Emerging Telecommunications
  Technologies}}, p. e2934, Mar. 2015.

\bibitem{doron1994}
M.~A. Doron and E.~Doron, ``Wavefield modeling and array processing. {{I}}.
  {{Spatial}} sampling,'' \emph{IEEE Transactions on Signal Processing},
  vol.~42, no.~10, pp. 2549--2559, 1994.

\bibitem{costa2012}
M.~Costa, A.~Richter, and V.~Koivunen, ``{{DoA}} and {{Polarization
  Estimation}} for {{Arbitrary Array Configurations}},'' \emph{IEEE
  Transactions on Signal Processing}, vol.~60, no.~5, pp. 2330--2343, May 2012.

\bibitem{proakis2008}
J.~G. Proakis and M.~Salehi, \emph{\BIBforeignlanguage{eng}{Digital
  {{Communications}}}}, 5th~ed.\hskip 1em plus 0.5em minus 0.4em\relax Boston,
  Mass.: {McGraw-Hill}, 2008.

\bibitem{kay1993}
S.~M. Kay, \emph{\BIBforeignlanguage{English}{Fundamentals of {{Statistical
  Signal Processing}}, {{Volume I}}: {{Estimation Theory}}}}, 1st~ed.\hskip 1em
  plus 0.5em minus 0.4em\relax Englewood Cliffs, N.J: {Prentice Hall}, Apr.
  1993.

\bibitem{olver2010}
F.~W.~J. Olver and {National Institute of Standards and Technology (U.S.)},
  Eds., \emph{{{NIST Handbook}} of {{Mathematical Functions}}}.\hskip 1em plus
  0.5em minus 0.4em\relax Cambridge; New York: {Cambridge University Press :
  NIST}, 2010.

\end{thebibliography}
	\bibliographystyle{IEEEtran}

\end{document}